\documentclass[12pt]{article}
\usepackage{amssymb}
\usepackage{amsfonts}
\usepackage{amsmath,amsthm}
\usepackage{graphics,epsfig}
\usepackage{subfigure}
\usepackage{graphicx,epsfig, color }
\oddsidemargin 10mm \numberwithin{equation}{section}
\def\be{\begin{equation}}
\def\ee{\end{equation}}
\begin{document}
\begin{center}
{{\bf {Modified Gauge Invariant Einstein Maxwell Gravity and
Stability of Spherical perfect fluid Stars with Magnetic
Monopoles}} \vskip 0.5 cm { Hossein Ghaffarnejad \footnote{E-mail
address: hghafarnejad@semnan.ac.ir} and ~~Leyla Naderi
\footnote{E-mail
address: l.naderi@semnan.ac.ir}}} \\
\vskip 0.1 cm {\textit{Faculty of Physics, Semnan University, P.C.
35131-19111, Semnan, Iran}
 } \vskip 0.1 cm
\end{center}
\begin{abstract}
As an alternative gravity model we consider an extended
Einstein-Maxwell gravity containing a gauge invariance property.
Extension is assumed to be addition of a directional coupling
between spatial electromagnetic fields with the Ricci tensor.  We
will see importance of the additional term in making a compact
stellar object and value of its radius. As an application of this
model we substitute ansatz of magnetic field of a hypothetical
magnetic monopole which has just time independent radial component
and for matter part we assume a perfect fluid stress tensor. To
obtain spherically symmetric internal metric of the perfect fluid
stellar compact object we solve Tolman-Oppenheimer-Volkoff
equation with a polytropic form of equation of state as
$p(\rho)=a\rho^2$. Using dynamical system approach we study
stability of the solutions for which arrow diagrams show saddle
(quasi stable) for $a<0$ (dark stars) and sink (stable) for $a>0$
(normal visible stars). We check also the energy conditions, speed
of sound and Harrison-Zeldovich-Novikov static stability criterion
for obtained solution and confirm that they make stable state.
\end{abstract}
\section{Introduction}
High energy astronomical compact objects in cosmic scales are
considered as excellent laboratories for investigating
astrophysical phenomena, and their relationship with nuclear and
elementary particles physics has opened a new approach to modern
astrophysics. High energy astronomical compact objects include for
instance, neutron stars, quark stars, boson stars, white dwarfs,
and black holes can be formed when a massive star runs out of its
fuel and therefore cannot remain stable against its own gravity
and then collapses \cite{1},\cite{2}. Depending on total value of
the mass of the star, the collapse changes the star`s
configuration and then initiates a new structure. In general a
star is stable when the pressure force from the gas atoms is equal
to its gravitational force and otherwise will be unstable. The
stability of the star can be investigated in the presence of both
electric and magnetic fields. Solving the Einstein-Maxwell field
equations for compact stars with the charged anisotropic fluid
model give more stable solutions than for neutral stars. The
presence of charges create repulsive forces against the
gravitational force, and so it causes to stable more for stars
with higher total mass and so larger redshift \cite{3}. In the
core of neutron stars, there is possibility of hadron-quark phase
transition. Charged quarks can create more stable quark stars than
neutron nuclei.  In theories beyond the standard model, the effect
of dark matter on the internal structure of the neutron stars
suggests that the neutron stars is mixed with dark matter in the
core and it is surrounded by a shell. This feature affects the
stellar mass-radius relation such that dark matter effects are
responsible for reducing the stellar mass, while the main effect
of the shell is to increase the stellar radius \cite{6}.
 In compact objects mixed with normal matter and dark matter, as the central pressure of dark matter increases,
 the neutron stars becomes unstable and the white dwarfs will have unusual masses and radii. Therefore, the resulting
 object will have unusually small mass and radius \cite{4}. When enough non-destructive dark matter accumulates on a neutron stars,
it creates a central degenerate star. If the mass of the dark
matter in the star reaches the Chandrasekhar mass limitation of
the star, the dark matter leads to collapse the mixed neutron
stars \cite{7}. The stability can also be investigated for compact
stars that are affected by strong magnetic fields and so affect
the process of stellar evolution. Surface Magnetic fields observed
in stars can be divided into two categories: the fossil and dynamo
hypothesis. The fossil hypothesis is used to explain magnetism in
massive stars, and the dynamo hypothesis, which is used for the
inner space of stars, shows the effects of a strong magnetic field
on the propagation of gravitational waves \cite{8}. Stability of
stars has provided via many gravitational models in which the main
question is whether a small perturbation can rapidly decay in
comparison to the model's parameters or not \cite{9}.
Hydrodynamical simulations in general theory of relativity have
been applied to investigate the dynamical stability of
differentially rotating neutron stars \cite{11}. Dynamical
instability of a star undergoing a dissipative collapse, has been
explored by considering the role of pressure anisotropy \cite{12}.
In general it is confirmed that the magnetic fields has a main
role in evolution and stability of the stars. For instance it is
obvious that sunspots are the largest concentration of complex
magnetic flux \cite{13}. Also there is inferred that energy source
of emission from magnetars is magnetic field (see for instance
\cite{14} and \cite{15}). Furthermore, the origin and dynamics of
magnetic fields in the surface of massive stars have been studied
in ref \cite{16}. Extensive study of the evolution of magnetic
field has been performed in rotating radiative zones of
intermediate-mass stars \cite{17}. Due to the importance of the
magnetic field effects on stability of stars we like to study
stability of a spherical perfect fluid stellar compact object in
presence of radial magnetic field of hypothetical magnetic
monopoles \cite{23},
\cite{24},\cite{25},\cite{26},\cite{27},\cite{28}. Pierre Curie
pointed out in 1894 \cite{29} that magnetic monopoles could
conceivably exist, despite not having been seen so far. From
quantum theory of matter, Paul Dirac \cite{30} showed that if any
magnetic monopoles exist in the universe, then all electric charge
in the universe must be quantized (Dirac quantization
condition)\cite{31}.  Since Dirac's paper, several systematic
monopole searches have
 been performed.
 Experiments in 1975 \cite{32} and 1982 \cite{33} produced candidate events that were initially interpreted \cite{32} as monopoles, but are now
 regarded inconclusive. Therefore, it remains an open question whether the magnetic monopoles exist. Further advances in theoretical
  particle physics, particularly developments in grand unified theories and quantum gravity, have led to more compelling arguments
   that monopoles do exist. Joseph Polchinski, provided an argument from string theory, confirming the existence of magnetic unipolarity,
   which has not yet been observed by experimental physics \cite{34}. These theories are not necessarily inconsistent with the experimental
    evidence. In some theoretical models, magnetic monopoles are unlikely to be observed, because they are too massive to create in particle
    accelerators, and also too rare in the Universe to enter a particle detector with much probability \cite{34}.\\
    It is well known that
     the ordinary matter consists of fermions in fact. But the fermions composed in such a way that
     the final products have integer spins. For real fermions
      as matter sources, we have to use spinors which can not be directly included in Einstein's GR equation.
      By the way, in cosmology, we consider the perfect fluid as a thermodynamics representation of the matter content
      of the universe and it is not constructed from some elementary particles. The energy density and pressure are two independent thermodynamics variables
   and they are related to each other via equation of state $p(\rho).$ However there is `Thomas-Fermi approximation` where two
assumptions are considered usually: (a) All gravitational and
other possible sources are slowly varying fields with respect to
fermion fields and so they do not interact with each other so that
one can use mean field theory (macroscopic quantities) instead of
dynamical microscopic fermion fields. (b) the fermion gas is at
equilibrium so that all the macroscopic quantities are time
independent and its stress tensor behaves as perfect fluid which
in the isotropic form is described by mass/energy density $\rho$
and hydrostatic isotropic pressure $p$ (see for instance
\cite{Lee} and \cite{Lor} for more details). On the other side
exotic $\alpha$ dependent term in our used lagrangian (see eq.
(\ref{action})) produces $\Theta$ term of stress tensor (\ref{25})
which does not satisfies covariant conservation condition (Bianchi
identity) alone and thus we need other matter stress tensor to
balance this inconsistency. Hence we assume the matter term to be
a perfect fluid stress tensor with polytropic type of equation of
state $p=a\rho^b.$ Usually the latter form of equation of state is
used for neutron stars with $b=2$. By using dynamical system
approach (see introduction section in \cite{37}) and solving TOV
equations we obtained parametric critical points in the phase
space and to check which of the internal metric solutions near the
critical points in phase space are physical we investigate null
energy condition (NEC), weak energy condition (WEC) and strong
energy condition (SEC) together with regularity, causality and
Harrison-Zeldovich-Novikov static stability (HZN) conditions.
 Layout of this paper is as follows.
\\
In section two we present proposed modified Einstein-Maxwell
gravity model together with its physical importance. As an
magnetic source to produce a spherically symmetric static metric
of an stellar object we consider radial magnetic field of a
magnetic monopole charge and derive  TOV equations of internal
metric of the system for stress tensor of perfect isotropic fluid.
We see that the equations of the fields are nonlinear and so we
must be use dynamical system approach to obtain solutions of the
fields near the critical points in phase space. This is done in
the section of third. Physical analysis of the obtained solutions
are dedicated to the section four.The last part of this work is
devoted to conclusions and prospects for the development of the
work. \section{Gravity model} Let us start with the following
generalized Einstein Maxwell gravity.
\begin{equation}\label{action}I=\int d^4x\sqrt{g}\left[\frac{R}{16\pi}+\frac{1}{4}F_{\mu\nu}F^{\mu\nu} +\alpha
F_{\rho\mu}R^{\mu}_{\eta}F^{\eta\rho} \right]+I_{matter},
\end{equation} where $g$ is absolute value of determinant of the metric field and dimensions in the coupling constant $\alpha$\footnote{Usually,
 the $\alpha$ exotic term in the above lagrangian is so called the non-minimal susceptibility
 tensor \cite{Alex} and in extended version it can be defined by the Reimann and Weyl tensors. } is
square of length and we write the action in the geometric units
$c=G=1$. $R_{\mu\nu}(R)$ is Ricci tensor (scalar) and
anti-symmetric electromagnetic tensor field $F_{\mu\nu}$  is
defined by $F_{\mu\nu}=\nabla_\mu A_\nu-\nabla_\nu A_\mu$ which
for torsion free Riemannian geometries can be rewritten as
that\begin{equation}\label{22} F_{\mu\nu}=\partial_\mu
A_\nu-\partial_\nu A_\mu.
\end{equation} In fact motivation of such a model is given previously in
the paper \cite{turner} to investigate how is broken conformal
invariance symmetry of the electromagnetic field in the
cosmological context.
 This is needed to produce large scale magnetic fields $(\sim Mpc)$ with high intensity. Regretfully, a pure $U(1)$ gauge theory with
 the standard Lagrangian $F_{\mu\nu}F^{\mu\nu}/4$ is conformal invariant and so for a Robertson-Walker spacetime with scale factor $a(t),$
 the magnetic field intensity decreases as $1/a^2$ and in the de Sitter inflationary epoch is ineffective and the vacuum energy density is
  dominated. While, today, it is obvious that magnetic fields are present throughout the universe and it plays an important role in
  astrophysical situations. For instance presence of high intensity magnetic field generates a high pressure which prevents the star from
   contracting. Even it is necessary
   to initiate substantial currents in superconducting cosmic strings which can be possible just by presence of high intensity cosmic
    magnetic fields (see  \cite{turner} and reference therein).To do so we must add some additional suitable scalars to the
     Lagrangian $F_{\mu\nu}F^{\mu\nu}/4$ such as given in the equation (\ref{action}).
It is easy to check that the above action functional is not
changed by transforming $A_{\mu} \to A_{\mu}+\partial_{\mu} \xi$
where $A_{\mu}$ is four vector electromagnetic potential field and
$\xi$ is a scalar gauge field, because by using the mentioned
transformation, we obtain $F_{\mu\nu}\to F_{\mu\nu}$. Varying the
above action with respect to the metric tensor field $g_{\mu\nu}$
reads the Einstein metric field equations such that
\begin{equation}\label{22}
G_{\mu\nu}=8\pi
T_{\mu\nu}^{total}=8\pi[T_{\mu\nu}^{EM}+\alpha\Theta_{\mu\nu}+T_{\mu\nu}^{(matter)}],\end{equation}
where
\begin{equation}\label{23}
G_{\mu\nu}=R_{\mu\nu}-\frac{1}{2}g_{\mu\nu}R,
\end{equation}
is the Einstein tensor defined by the Ricci tensor $R_{\mu\nu}$
and the Ricci scalar $R=g^{\mu\nu}R_{\mu\nu}$,
\begin{equation}\label{24}
T^{EM}_{\mu\nu}=-\frac{1}{8}\left[F_{\mu\alpha}F^\alpha_\nu+F_{\beta\nu}F^\beta_\mu
-\frac{1}{2}g_{\mu\nu}F_{\alpha\beta}F^{\alpha\beta}\right]
\end{equation}
is traceless electromagnetic field stress tensor,
$$\Theta_{\mu\nu}=\frac{1}{4}g_{\mu\nu} F_{\rho\alpha} R^\alpha_\eta F^{\eta\rho}-\frac{1}{2}[
F_{\rho\mu} R_{\nu\eta}F^{\eta\rho}+F_{\rho\eta} R^\eta_\mu
F^\rho_\nu +F_{\rho\mu} R^\rho_\eta F^\eta_\nu] $$
$$+\frac{1}{4\sqrt{g}}\partial_\alpha\left[\partial_\eta\left(\sqrt{g}F^\alpha_\rho F^{\eta\rho}\right)\right]g_{\mu\nu}
-\frac{1}{2\sqrt{g}}\partial_\mu\left(\sqrt{g}F^\alpha_\rho
F^{\eta\rho}\right)\Gamma_{\nu\alpha\eta}$$$$
-\frac{1}{8\sqrt{g}}g_{\eta\mu}g_{\sigma\nu}\partial_\lambda\left[\partial_\alpha\left(\sqrt{g}F^\lambda_\rho
F^{\eta\rho}\right)g^{\alpha\sigma}\right]$$
\begin{equation}\label{25}
-\frac{1}{8\sqrt{g}}g_{\sigma\mu}g_{\lambda\nu}\partial_\eta\left[\partial_\alpha\left(\sqrt{g}F^\lambda_\rho
F^{\eta\rho}\right)g^{\alpha\sigma}\right]+\frac{1}{8\sqrt{g}}g_{\lambda\mu}g_{\eta\nu}\partial_\sigma\left[\partial_\alpha\left(\sqrt{g}
F^\lambda_\rho F^{\eta\rho}\right)g^{\alpha\sigma}\right]
\end{equation}
is gravity-photon interaction stress tensor and
$T_{\mu\nu}^{(matter)}$ is matter part stress tensor respectively.
Here we choose matter content of the system to be isotropic
perfect fluid with stress tensor \cite{GL}
\begin{equation}\label{mat}(T^{matter})^\mu_\nu=diag(-\rho,p,p,p)\end{equation}
where pressure is related to the density via a suitable equation
of state $p(\rho)=a\rho^b.$ Electromagnetic Maxwell field equation
is given by varying the action functional (\ref{action}) with
respect to the gauge field $A_{\mu}$ such that
\begin{equation}\label{26}
\nabla_{\nu}F^{\mu\nu}=2\alpha J^{\mu},
\end{equation}
where the  four current density, $J^{\mu}$, is defined by
\begin{equation}\label{27}
J^{\mu}=\nabla_\lambda C^{\lambda\mu}
\end{equation} in which $C^{\lambda\mu}$ is anti-symmetric tensor \begin{equation}C^{\lambda\mu}=-C^{\mu\lambda}
=\left(R^\mu_\eta F^{\eta\lambda}-
R^\lambda_\eta F^{\eta\mu}\right).\end{equation} One can show that
the Maxwell equation (\ref{26}) can be rewritten as follows.
\begin{equation}\label{Max}\nabla_\nu\tilde{F}^{\mu\nu}=0,~~~\tilde{F}^{\mu\nu}=F^{\mu\nu}+2\alpha C^{\mu\nu}\end{equation} and
for arbitrary anti-symmetric tensor $O^{\mu\nu}$ we have
\begin{equation}\nabla_\mu O^{\mu\nu}=\frac{\partial_\nu
(\sqrt{g}O^{\mu\nu})}{\sqrt{g}}\end{equation} in torsion free
curved spacetimes. In the differential geometry formalism of the
electromagnetic field, the above antisymmetric Faraday tensor can
be written as follows
\cite{35}.\begin{equation}\label{FF}\textbf{F}=\frac{1}{2}F_{\mu\nu}dx^\mu\wedge
dx^\nu=\textbf{B}+\textbf{E}\wedge
dx^0=d\textbf{A}+\textbf{E}\wedge dx^0
\end{equation}
in which \begin{equation}\textbf{E}=E_idx^i\end{equation} is
1-form electric field and
\begin{equation}\textbf{B}=\frac{1}{2}\epsilon_{ijk} B^idx^j\wedge dx^k\end{equation} is 2-form magnetic field.
 In fact they are spatial vector fields
and $i,j=1,2,3$ correspond to spatial coordinates while $x^0$
denotes to time coordinate in the curved background spacetime. In
the above equation $\epsilon_{ijk}$ is third rank totally
antisymmetric Levi Civita tensor density. Its numeric value is
$+1(-1)$ for $\{i,j,k\}=\{1,2,3\}$ and for any even (odd)
permutations while it takes zero value for any two repeated
indices. The equation (\ref{FF}) can be rewritten to the following
form also \cite{12}.
\begin{equation}\label{2333}F_{\mu\nu}=n_\mu E_{\nu}-n_{\nu}E_\mu+\epsilon_{\mu\nu\eta\lambda}B^\eta n^\lambda\end{equation}
where $n_{\mu}$ is a unit time-like vector field and is normal to
the spatial 3D hypersurface $x^0=const$ and so can be defined by
$n^\mu=-\nabla_{\mu}x^0/||\nabla_{\mu}x^0||.$ Consequently the
electric and magnetic fields components $\{E_i,B_{i}\}$ are
measured by a normal observer aligned to $n_{\mu}$ and so they are
absolutely spatial vector fields $E_\mu n^\mu=0=B_{\mu}n^\mu.$
 From ADM formalism in the $1+3$ decomposition of any 4D curved background spacetime metric the whole of spacetime can be
 foliated into hypersurfaces with constant time coordinate $x_0$ where $h_{ij}=g_{ij}$ are spatial 3-metric defined on the spacelike
 hypersurfaces.
In the other words the general form of line element
$ds^2=g_{\mu\nu}dx^\mu dx^\nu$ reads
\begin{equation}\label{ds}ds^2=-\alpha^2dt^2+h_{ij}(dx^i+\beta^i dt)(dx^j+\beta^jdt)\end{equation} in which  $\alpha$ is lapse  function and
$\beta^i$ is shift vector. In the definition (\ref{2333})
$\epsilon_{\mu\nu\eta\lambda}$ is fourth rank totally
antisymmetric Levi Civita tensor density. Its numeric value is
$+1(-1)$ for $\{\mu,\nu,\eta,\lambda\}=\{0,1,2,3\}$ and for any
even (odd) permutations while it takes zero value for any two
repeated indices. For time independent static curved spacetimes
the line element (\ref{ds}) takes a simpler forme because
$\beta^i=0$ and $\alpha$ and $h_{ij}$ take on just spatial
coordinates $x^i$. In this case we can apply a suitable
coordinates transformation to remove all non-diagonal components
of $h_{ij}$ such that
\begin{equation}\label{ds2}ds^2=-(\alpha dt)^2+(\gamma_idq^i)^2\end{equation}
in which $d\ell_i=\gamma_idq_i$ has length dimension and $dq^i$ is
spatial coordinates used in a local curvilinear frame in the 3D
space \cite{36}. For the line element (\ref{ds2}) the identity
(\ref{FF}) reads
\begin{equation}\label{Fij}F_{it}=\sqrt{\alpha}\gamma_i E_i,~~~F_{jk}=\epsilon_{ijk}B_i\gamma_{j}\gamma_{k}\end{equation}
in which repeated indexes for $\epsilon_{ijk}$ do not follow the
Einstein summation rule and just follows permutation cycles. In
the next section we write  the Einstein equations and the
Tolman-Oppenheimer-Volkoff equation for internal metric of a
spherically symmetric  static compact stellar object in presence
of magnetic  field of a magnetic monopole charge and stress tensor
of a perfect fluid with density $\rho$ and  pressure $p$ with
polytropic form of equation of state $p=a\rho^2.$ This form of
equation of state is used usually for Neutron stars \cite{Lei}.
\section{Tolman-Oppenheimer-Volkoff
equation} Line element for a general spherically symmetric curved
spacetime is given by
\begin{equation}\label{23}ds^2=-X(t,r)dt^2+Y(t,r)dr^2+r^2(d\theta^2+\sin^2\theta
d\varphi^2)\end{equation}  for which the equations (\ref{Fij})
reads \begin{equation}\label{233}F_{\mu\nu}
=\left(%
\begin{array}{cccc}
0 &-\sqrt{XY}E_r & -\sqrt{X}rE_\theta & -\sqrt{X}r\sin\theta E_\varphi \\
\sqrt{XY}E_r& 0 & \sqrt{Y}rB_\varphi& -\sqrt{Y}r\sin\theta B_{\theta} \\
\sqrt{X}rE_\theta & -\sqrt{Y}rB_\varphi & 0 & r^2\sin\theta B_r \\
\sqrt{X}r\sin\theta E_\varphi & \sqrt{Y}r\sin\theta B_\theta & -r^2\sin\theta B_r& 0 \\
\end{array}%
\right)
\end{equation}
and the $t,r$ components of the Einstein equation (\ref{22}) reads
\begin{equation}\label{e1}\frac{X^{\prime}}{X}=8\pi rY(T_{total})^r_r+\frac{(Y-1)}{2r},
\end{equation}
\begin{equation}\label{e2}\frac{Y^\prime}{Y}=-8\pi r Y(T_{total})^t_t-\frac{(Y-1)}{2r}
\end{equation}
and \begin{equation}\label{e3}\frac{\dot{Y}}{Y}=8\pi
rX(T_{total})^t_r\end{equation} where $^\prime$ and $\dot{~}$ are
partial derivatives with respect to $r$ and $t$ respectively. In
usual way to study internal metric of spherically symmetric object
$\theta\theta$ and $\varphi\varphi$ components of the Einstein
equations are not used and instead of them, one usually use the
Bianchi identity or equivalently, the covariant conservation
equation of matter stress tensor.  Covariant conservation equation
for the perfect fluid stress tensor (\ref{mat}) namely
$\nabla_{\mu}(T_{fermions})^\mu_\nu=0$ gives us equation of
motions for $\rho(t,r)$ and $p(t,r)$ such that
\begin{equation}\label{dotrho}\dot{\rho}=-\frac{\dot{Y}}{2Y}(\rho+p)\end{equation} and
\begin{equation}\label{prime}p^\prime=-\frac{X^\prime}{2X}(\rho+p).\end{equation}  It is easy to check that the only magnetic field
 having property of spherical symmetry is corresponded just to the magnetic monopole with assumed charge $q_m$ whose the magnetic potential is
\begin{equation}\label{212}A_{\varphi}(\theta)=-q_m\cos\theta.\end{equation} By regarding (\ref{22}) one can show that the
corresponding non-vanishing component of the Maxwell tensor field
for (\ref{212}) is \begin{equation}\label{25}
F_{\theta\varphi}=\partial_{\theta}A_{\varphi}=q_m\sin\theta
\end{equation} which by substituting into (\ref{233}) we obtain\begin{equation}
B_r(r)=\frac{q_m}{r^2}.\end{equation} This is similar to radial
electric field of an electric monopole charge
$E_r(r)=\frac{q_e}{r^2}.$  One can show that for the magnetic
monopole field (\ref{25}) we will have
\begin{equation}\label{EM}(T^{EM})^\mu_\nu=\frac{q_m^2}{8r^4}\left(%
\begin{array}{cccc}
  +1 & 0 & 0 & 0 \\
  0 & +1 & 0 &0 \\
  0 & 0 & -1 & 0 \\
  0 & 0 & 0 & -1 \\
\end{array}%
\right)\end{equation} and for $\Theta^\mu_\nu$ tensor we obtain
\begin{equation}\label{thet1}\Theta^t_r=-\frac{q_m^2}{2r^5}\bigg(\frac{\dot{X}}{X}+\frac{\dot{Y}}{Y}\bigg),
\end{equation}\begin{equation}\label{thet2}\Theta^t_t=\frac{q_m^2}{4r^6Y}\bigg[2(1-2Y)+r\bigg(\frac{X^\prime}{X}-\frac{Y^\prime}{Y}
\bigg)\bigg]\end{equation}\begin{equation}\label{thet3}\Theta^r_r=\frac{q_m^2}{r^6Y}\bigg[\frac{r}{2}\frac{X^\prime}{X}-(1+Y)\bigg]\end{equation}
and
\begin{equation}
\label{thet4}\Theta^\theta_\theta=\Theta^\varphi_\varphi=\frac{q_m^2}{8r^6Y}\bigg[7Y-8+4r\bigg(\frac{2Y^\prime}{Y}-
\frac{X^\prime}{X}
\bigg)+\frac{r^2}{2}\bigg(\frac{X^{\prime\prime}}{X}+\frac{Y^{\prime\prime}}{Y}\bigg)\end{equation}
$$+\frac{r^2}{4}\bigg(\frac{4X^\prime Y^\prime}{XY}+\frac{X^{\prime2}}{X^2}-\frac{Y^{\prime2}}{Y^2}
\bigg)+\frac{r^2Y}{2X}\bigg[\frac{2\dot{X}^2}{X^2}+\frac{\dot{Y}^2}{Y^2}+\frac{\dot{X}\dot{Y}}{XY}-\frac{\ddot{X}}{X}-\frac{\ddot{Y}}{Y}
\bigg]\bigg] .$$ It is easy to check that the magnetic monopole
field (\ref{25}) satisfies the Maxwell equation (\ref{Max}) as
trivially. By substituting (\ref{mat}), (\ref{EM}), (\ref{thet1}),
(\ref{thet2}) and (\ref{thet3})  into the equations (\ref{e1}),
(\ref{e2}) and (\ref{e3}) we obtain
\begin{equation}\label{p}p=-\frac{1}{16\pi r^2}-\frac{q_m^2}{8r^4}+
\frac{\alpha q_m^2}{r^6}+\frac{1}{16\pi
r^2Y}\bigg(1+\frac{16\pi\alpha q_m^2}{r^4}
\bigg)\end{equation}$$+\frac{1}{8\pi rY}\bigg(1-\frac{4\pi\alpha
q_m^2}{r^4}\bigg)\frac{X^\prime}{X}$$
\begin{equation}\label{rho} \rho=-\frac{1}{16\pi
r^2}-\frac{q_m^2}{8r^4}+\frac{1}{16\pi
r^2Y}\bigg(1-\frac{8\pi\alpha q_m^2}{r^4} \bigg)-\frac{\alpha
q_m^2}{4r^5Y}\frac{X^\prime}{X}\end{equation}
$$-\frac{1}{8\pi
rY}\bigg(1-\frac{2\pi\alpha q_m^2}{r^4}\bigg)\frac{Y^\prime}{Y}$$
and
\begin{equation}\label{tr}\dot{X}+\bigg(X+\frac{r^4}{4\pi\alpha q_m^2}\bigg)\frac{\dot{Y}}{Y}=0.\end{equation}
$\theta,\varphi$ components of the Einstein equations (\ref{22})
have  similar form and they are a constraint condition between the
metric solutions $X(r)$ and $Y(r).$ To investigate internal metric
of stellar compact object we use the covariant conservation
equation of matter stress tensor given by (\ref{dotrho}) and
(\ref{prime}) instead of the $\theta,\varphi$ components of the
Einstein equation. To solve these equations we need also an extra
relation between the pressure and the density $p=f(\rho)$ called
as equation of state. In this paper we use general form of
polytropic equation of state
\begin{equation}\label{pr}p(\rho)=a\rho^b\end{equation} where dimensionless parameter $b$ is
 called as the constant adiabatic exponent but dimensional parameter $a$ is called as barotropic index. This
kind of equation of state is usually applicable for relativistic
stares for instance the neutron stars ($b=2,$\cite{Lei}), the
boson or the fermion stars ($b=\frac{4}{3}$, \cite{7}). We see in
the subsequent sections that $a>0 (a<0)$ corresponds to visible
(dark) stars with stable (quasi stable) nature.
 We are now
in position to solve the above dynamical equations as follows.\\
 To study stability condition of
the Einstein metric solutions it is convenient we consider static
time-independent version of the line element (\ref{23}) which is
dependent just to $r$ coordinate. Hence we ignore all partial time
derivatives of the fields. In this case one can see that
(\ref{dotrho}) and (\ref{tr}) are removed trivially, while
(\ref{prime}) by substituting the equation of state (\ref{pr})
reads
\begin{equation}\label{X}X(r)=\frac{K}{(1+a\rho^{b-1})^{\frac{2b}{b-1}}}\end{equation}
in which $K$ is a suitable integral constant. By substituting
(\ref{X}) and the equation of state (\ref{pr}), the equations
(\ref{p}) and (\ref{rho}) read to the following forms
respectively:
\begin{align}\label{rhop}\rho^\prime&=\bigg(\frac{1+a\rho^{b-1}}{ab\rho^{b-2}}\bigg)\frac{8\pi
r}{\big(1-\frac{4\pi\alpha
q_m^2}{r^4}\big)}\bigg[Y\bigg(a\rho^b+\frac{1}{16\pi
r^2}+\frac{q_m^2}{8r^4}-\frac{\alpha
q_m^2}{r^6}\bigg)\notag\\&-\frac{1}{16\pi
r^2}\bigg(1+\frac{16\pi\alpha q_m^2}{r^4}\bigg)\bigg]\end{align}
and
\begin{align}Y^\prime&=\bigg(1-\frac{4\pi\alpha q_m^2}{r^4}\bigg)^{-1}\bigg(1-\frac{2\pi\alpha q_m^2}{r^4}\bigg)^{-1}\bigg\{
Y\bigg[\frac{\pi\alpha q_m^2}{r^5}\bigg(1+\frac{16\pi\alpha q_m^2}{r^4}
\bigg)\notag\\&-\frac{1}{2r}\bigg(1-\frac{4\pi\alpha
q_m^2}{r^4}\bigg)\bigg(1-\frac{8\pi\alpha
q_m^2}{r^4}\bigg)\bigg]-Y^2\bigg[\frac{\pi\alpha
q_m^2}{r^3}\bigg(16\pi a\rho^b+\frac{1}{r^2}\notag\\& +\frac{2\pi
q_m^2}{r^4}-\frac{16\pi\alpha q_m^2}{r^6}\bigg)+4\pi
r\bigg(1-\frac{4\pi\alpha q_m^2}{r^4}\bigg)\bigg(16\pi
\rho+\frac{1}{r^2}+\frac{2\pi
q_m^2}{r^4}\bigg)\bigg]\bigg\}.\end{align} These equations show a
two dimensional phase space $\{Y,\rho\}$ which can be solved near
the critical points by approach of dynamical systems. We know that
density and pressure of a compact stellar object should vanishes
on its surface. The equation (\ref{rhop}) is singular at $\rho=0$
for $b\neq2$ and so we substitute ansatz $b=2$ in that equation.
In this case we can assume that the critical radius of the compact
object is its radius $R=r_c$ if it satisfies the critical point
equations $\rho^\prime=0=Y^\prime$ for which
\begin{align}\label{crit}&\rho_c(R)=0,~~~r_c=R\notag\\&
Y_c=\frac{1+\frac{16\pi\alpha q_m^2}{r^4_c}}{1+\frac{2\pi
q_m^2}{r_c^2}-\frac{16\pi\alpha q_m^2}{r_c^4}}\end{align} and
critical radius of the stellar compact object $r_c$ is obtained by
the equation
\begin{align}\label{rc}\bigg(1-\frac{4\pi\alpha q_m^2}{r^4_c}\bigg)&\bigg[
\bigg(1-\frac{8\pi\alpha q_m^2}{r^4_c}\bigg)\bigg(1+\frac{2\pi
q_m^2}{r_c^2}-\frac{16\pi\alpha
q_m^2}{r^4_c}\bigg)\notag\\&+8\bigg(1+\frac{16 \pi\alpha
q_m^2}{r^4_c}\bigg)\bigg(1+\frac{2\pi\alpha
q_m^2}{r^4_c}\bigg)\bigg]=0.\end{align} The first term in the
above equation has unacceptable solution as $r_c=(4\pi\alpha
q_m^2)^\frac{1}{4}$ because, coefficients of the Jacobi matrix
calculated at the belove diverge to infinity. Hence we exclude
this solution from the physical critical radius. Other physical
solutions of the critical radiuses are obtained from the second
part of the  equation (\ref{rc}). However, one can obtain Jacobi
matrix components as
\begin{equation}J_{ij}=\frac{\partial O^\prime_i}{\partial O_j}\bigg|_{\rho_c=0,Y=Y_c,b=2}=\left(%
\begin{array}{cc}
  0 & J_{12} \\
  J_{21} & J_{22} \\
\end{array}%
\right)\end{equation} in which
\begin{align}&J_{12}=\frac{1}{4ar_c}\bigg(1-\frac{4\pi\alpha q_m^2}{r_c^4}\bigg)^{-1}\bigg[1+\frac{2\pi
q_m^2}{r_c^2}-\frac{16\pi\alpha q_m^2}{r_c^4}\bigg]\notag\\&
J_{21}=\frac{-64\pi^2 r_c\big(1+\frac{16\pi\alpha
q_m^2}{r_c^4}\big)^2}{\big(1-\frac{2\pi\alpha
q_m^2}{r_c^4}\big)\big(1+\frac{2\pi
q_m^2}{r_c^2}-\frac{16\pi\alpha q_m^2}{r_c^4}\big)^2}\notag\\&
J_{22}=\frac{-\frac{\pi\alpha
q_m^2}{r_c^5}\big(1+\frac{16\pi\alpha
q_m^2}{r_c^4}\big)}{\big(1-\frac{2\pi\alpha
q_m^2}{r_c^4}\big)\big(1-\frac{4\pi\alpha
q_m^2}{r_c^4}\big)}-\frac{\frac{1}{2r_c}\big(1-\frac{8\pi\alpha
q_m^2}{r_c^4}\big)}{\big(1-\frac{2\pi\alpha
q_m^2}{r_c^4}\big)}\notag\\&
-\frac{\frac{8\pi}{r_c}\big(1+\frac{2\pi
q_m^2}{r_c^2}\big)\big(1+\frac{16\pi\alpha
q_m^2}{r_c^4}\big)}{\big(1-\frac{2\pi
q_m^2}{r_c^2}\big)\big(1+\frac{2\pi
q_m^2}{r_c^2}-\frac{16\pi\alpha q_m^2}{r_c^4}\big)}.\end{align} In
the dynamical system approach we can now obtain solutions of the
field equations near the critical point (\ref{crit}) by
\begin{equation}\label{arrow}\frac{d}{dr}\left(%
\begin{array}{c}
  \rho \\
  Y \\
\end{array}%
\right)=\left(%
\begin{array}{cc}
  0 & J_{12} \\
  J_{21} & J_{22} \\
\end{array}%
\right)\left(%
\begin{array}{c}
  \rho \\
  Y \\
\end{array}%
\right)\end{equation} which reads to the following
equations:\begin{align}\label{rhor}&\rho(r)=J_{12}\int_{r<r_c}^{r_c}Y(r)dr\notag\\&
Y^{\prime\prime}-J_{22}Y^{\prime}-J_{12}J_{21}Y=0.
\end{align} To
solve these equations we must use the initial conditions given by
the critical point (\ref{crit}) such that
\begin{equation}\label{Yr}Y(r)=Y_c\bigg[\frac{\omega_- e^{\omega_+(r-r_c)}-\omega_+ e^{\omega_-(r-r_c)}}{\omega_--\omega_+}\bigg]\end{equation}
in which
\begin{equation}\label{omega}\omega_{\pm}=\frac{J_{22}\pm\sqrt{J^2_{22}+4J_{12}J_{21}}}{2}.\end{equation}
By substituting the metric solution (\ref{Yr}) into the equation
(\ref{rhor}) and calculation of its integration one find
\begin{equation}\label{rhosol}\rho(r)=\frac{Y_cJ_{12}}{\omega_--\omega_+}\bigg[\frac{\omega_-}{\omega_+}(1-e^{\omega_+(r-r_c)})-\frac{\omega_+}{\omega_-}(1-e^{\omega_-(r-r_c)})
\bigg].\end{equation} In fact $\omega_{\pm}$ given by the equation
(\ref{omega}) is obtained from the secular equation of the Jacobi
matrix $det(J_{ij}-\omega\delta_{ij})=0.$ In the dynamical system
approach the obtained solutions near the critical points are
stable if the eigenvalues $\omega_{\pm}$ have negative values when
they are real and when they are complex numbers then their real
part should be negative. In the cases with positive values for
real eigenvalues the obtained solutions are not stable. Hence we
extract the choices with negative values for real part of
$\omega_{\pm}.$ Also we can obtain exactly numeric values for the
critical points $r_c$ given by the equation (\ref{crit}) but it is
useful we study asymptotic behavior of the obtained solutions for
large radiuses $r_c>>>|q_m|.$ In this case the equation
(\ref{crit}) reads
\begin{equation}9\bigg(\frac{r_c}{q_m}\bigg)^4+2\pi\bigg(\frac{r_c}{q_m}\bigg)^2+120\pi\alpha q_m^2\approx0\end{equation} with solutions \begin{equation}1<<
\bigg(\frac{r_c}{q_m}\bigg)^2= \frac{\sqrt{\pi^2-1080\pi\alpha
q_m^2}-\pi}{9}\approx\bigg(-\frac{40}{3}\alpha
q_m^2\bigg)^\frac{1}{2},~~~\alpha<0\end{equation} or
\begin{equation}r_c\approx q_m\bigg(-\frac{40}{3}\alpha q_m^2\bigg)^\frac{1}{4},~~~\alpha<0.\end{equation} For $r_c>>q_m$ one can show
\begin{align}&\lim_{\frac{r_c}{q_m}\to\infty} J_{12}\sim\frac{1}{4a r_c},~~~\lim_{\frac{r_c}{q_m}\to\infty}J_{21}\sim-64\pi^2 r_c\notag\\&
\lim_{\frac{r_c}{q_m}\to\infty}J_{22}\sim0,~~~\lim_{\frac{r_c}{q_m}\to\infty}Y_c(r)\sim1,~~~\lim_{\frac{r_c}{q_m}\to\infty}\omega_{\pm}\sim\frac{\pm4\pi
i}{\sqrt{a}}.\end{align} By substituting these asymptotic behavior
of the parameters into the  solutions (\ref{Yr}) and
(\ref{rhosol}) we find \begin{equation}\label{sol}Y(r)\approx
\cos[\Omega(1-\bar{r})],~~~\bar{\rho}(\bar{r})=\frac{\rho(r)}{\rho(0)}\approx\frac{\sin[\Omega(1-\bar{r})]}{\sin[\Omega]},~~~0\leq
\bar{r}\leq1\end{equation} in which
\begin{equation}\Omega=\frac{4\pi r_c}{\sqrt{a}},~~~\bar{r}=\frac{r}{r_c}\end{equation} and we defined central density as
\begin{equation}\rho(0)=\frac{\sin[\Omega]}{4a\Omega},\end{equation} For this  density function, one find mass function such that
\begin{equation}M(r_c)=\int_0^{r_c}\rho(r)dr=\frac{\rho(0)\sqrt{a}r_c^2}{\sin[\Omega]}\bigg[1-\bigg(\frac{\sin\Omega}{\Omega}\bigg)^2
\bigg]\end{equation} for which
\begin{equation}\frac{2M}{R}=\frac{2M(r_c)}{r_c}=\frac{1}{8\pi}\bigg[1-\bigg(\frac{\sin\Omega}{\Omega}\bigg)^2
\bigg]<1.\end{equation} This result shows that our obtained
solutions describe a regular star without a Schwarzschild-like
horizon.  To see stability of the solution it is useful to plot
arrow diagrams of the dynamical equations given by (\ref{arrow})
such that
\begin{equation}\dot{\bar{\rho}}\approx
Y,~~~\dot{Y}\approx \epsilon \bar{\rho},~~~\bar{\rho}=4\pi
a\rho,~~~\dot{~}=\frac{1}{r_c}\frac{d}{d\tau},~~~r=r_c\tau,~~~\epsilon=-\frac{16r_c^2}{a}.\end{equation}
See figure 1 which is plotted for ansatz $\epsilon=1$ and
$\epsilon=-1.$
 To
be more sure of the obtained solutions, we investigate on these
solutions some physical conditions that a real compact stellar
fluid must be had.
\section{Physical analysis of the metric solution}
A realistic
stellar model should satisfy some physical properties including
 the energy conditions, regularity, causality and stability. In this section we check all these properties for the obtained
 solutions.
\subsection{Energy conditions}
Energy conditions for a physical perfect fluid model are included
to three parts the so called null energy condition (NEC) with
$\rho\geq0,$ weak energy condition (WEC) with $\rho-p\geq0$ and
strong energy condition (SEC) with $\rho-3p\geq0.$ By looking at
the diagrams given in the figures 1-d, 2-a, 2-b, 2-c and 2-d one
can infer that NEC is dependent to value of the dimensionless
critical radius $\Omega.$ These diagrams show that by raising
$\Omega>3$ then, sign of the density function changes to negative
sign for regions $\bar{r}>0.2$ but for $\Omega\leq3$ we have
$\rho>0$ for full region $0<\bar{r}\leq1.$ To study WEC we
substitute $p=a\rho^2$ to obtain $\rho(1-a\rho)>0$ which reads to
the condition $a\rho<1.$ By substituting the obtained solution
(\ref{sol}) the WEC called as $a\rho<1$ reads
\begin{equation}\label{I}WEC:~~~\frac{\sin[\Omega(1-\bar{r})]}{4\Omega}\leq1\end{equation}
and for SEC called as $3a\rho\leq1$ we obtain same inequality
condition such that
\begin{equation}\label{II} SEC:~~~~\frac{3\sin[\Omega(1-\bar{r})]}{4\Omega}\leq1
\end{equation}
We plot diagrams of the above inequalities in figure 3.
  \subsection{Regularity} By looking at the obtained density function (\ref{sol}) one can infer that it is convergent
  regular function for
   $0\leq \bar{r}\leq1$. Furthermore arrow diagrams show that sink stable state for a regular
  visible stellar compact object with $a>0$
  while for $a<0$ which is so called as dark stars the solutions has quasi stable nature in the arrow diagram and so one can infer that our obtained solutions
  behave same as stellar compact object with normal
   (non-dark) matter with positive barotropic index $a>0.$
 \subsection{Casuality} The speed of sound
$v^2=\frac{dp}{d\rho}$ for a compact stellar object should be less
than the speed of light $c=1$ and so by substituting the equation
of state $p=a\rho^2$ one can obtain speed of sound for our model
as
\begin{equation}v=\sqrt{2a\rho}=\sqrt{\frac{\sin[\Omega(1-\bar{r})]}{2\Omega}}
\end{equation}
which its diagram is plotted vs $\bar{r}$ for different values of
the $\Omega$ parameter in figure 4-a.  By looking at this  diagram
one can infer that the case $\Omega=4$ is not physical because
does not satisfy the causality condition near the center
$0<\bar{r}<0.2.$ In other words  it is complex imaginary which is
not seen in the diagram while other cases $0<\Omega<3$ satisfy the
causality condition completely.
\subsection{Stability} One of ways to check gravitational
stability of a stellar system to be not collapsing is
investigation of numeric values of the adiabatic index of the
perfect fluid which in case of isotropic state is defined by
$\Gamma=\frac{dp}{d\rho}(1+\rho/p)$ \cite{vF}, \cite{qw} . When
$\Gamma\geq\frac{4}{3}$ then a stellar fluid object is said to be
stable from gravitational collapse. For our model one can show
that
\begin{equation}\Gamma=2(1+a\rho)\geq\frac{4}{3}\end{equation}
which means that our obtained solutions is free of gravitational
collapse just for $a\rho\geq-\frac{1}{3}$ such that
\begin{equation}\frac{\sin[\Omega(1-\bar{r})]}{4\Omega}\geq-\frac{1}{3}.\end{equation} We plot diagram of this
inequality for different values of the parameter $\Omega$ vs
$0\leq\bar{r}\leq1 $ in figure 4-b.
 Other way to study stability of a compact gaseous
stellar object in presence of radial perturbations was provided at
a first time by Chandrashekhar (see \cite{Chan1} and
\cite{Chan2}). It was developed and simplified by Harrison et al
\cite{Har} and Zeldovich with collaboration of Novikov \cite{Zel}.
This is now well known as `Harrison-Zeldovich-Novikov (HZN) static
stability criterion` which infers that any solution describes
static and stable (unstable) stellar structure if  the
gravitational total mass $M(\rho(0))$ is an increasing
(decreasing) function versus the central density $\rho(0)$ i.e,
$\frac{\partial M}{\partial \rho(0)}>0 (<0)$ under radial
pulsations. For our model the HZN condition reads
\begin{equation}HZN=\frac{1}{\sqrt{a}r_c^2}\frac{\partial M(\rho(0))}{\partial\rho(0)}=\frac{1}{\sin\Omega}\bigg[1-\bigg(\frac{\sin\Omega}{\Omega}\bigg)^2\bigg]\end{equation}
which we plot its diagram vs $\Omega$ in figure 4-c. It shows
stability condition for choices $0<\Omega<\pi$ which obey the
other diagrams given by figures 2. To see this one can look
behavior of the red-dash-lines in figures 1-d, 2-a,2-b-2 and 2-c
where the density functions take on positive values (NEC) for full
interior region of the compact stellar object $0<\bar{r}<1$ but
not for $\Omega=4$ given by the figure 2-d.
\section{Conclusion}
In this paper we considered a modified Einstein-Maxwell gravity
where modification is the directional dependence of coupling
between the electromagnetic field and Ricci tensor. Motivation of
this kind of extensions is support of cosmic inflation with cosmic
magnetic fields instead of unknown dark sector of the
matter/energy. Hence we encouraged to investigate such a model for
a stellar compact object system with a perfect fluid kind of
matter source. To consider magnetic field of the model we use
ansatz of magnetic field of magnetic monopole charge. We solved
Tolman-Oppenheimer-Volkoff equation for interior metric of a
spherically symmetric static perfect fluid. We used dynamical
system approach to do because of nonlinearity form of the
dynamical equations and obtained solutions of the fields near
critical points. Our obtained solutions are physical because they
satisfy energy conditions (NEC, WEC, SEC) and also the
Harrison-Zeldovich-Novikov static stability. Also we check that
sound speed is less than the light velocity and the obtained
solutions obey the causality. In this work we use mean field
theory approximation for matter stress tensor with mean energy
density and isotropic pressure and we do not consider microscopic
behavior of the matter source. As an extension of this work we
like to study in our next work, effects of anisotropic imperfect
fluid from point of view of its microscopic behavior in presence
of magnetic monopole field.
\\
\\
\textbf{Acknowledgement}
\\
\\
This work was supported in part by the Semnan University Grant No.
1678-2021 for Scientific Research
\section{Data Availability Statement} No Data associated in the manuscript

\begin{figure}
\centering\subfigure[{}]{\label{11011}
\includegraphics[width=0.45\textwidth]{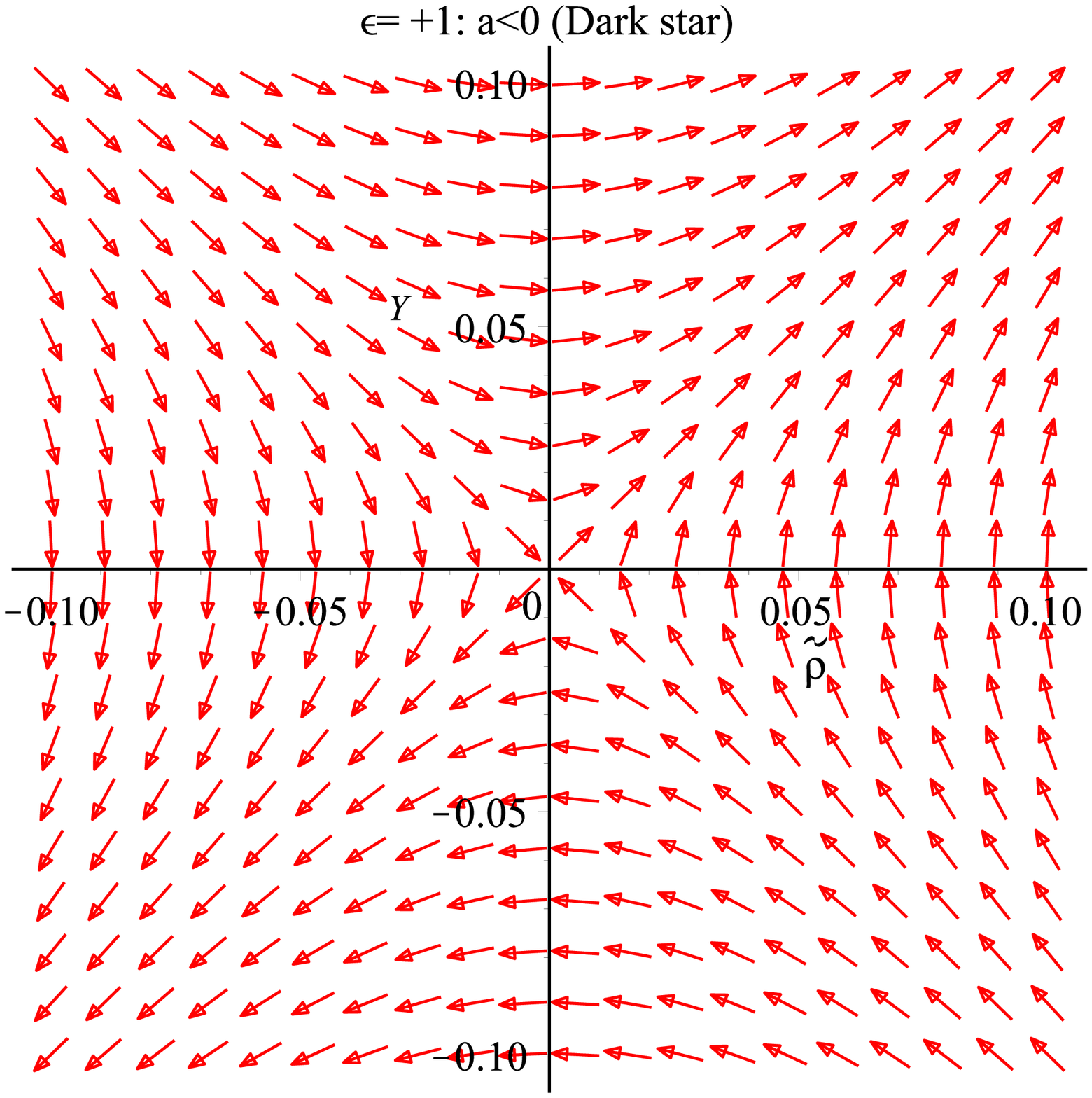}}
\hspace{2mm}\subfigure[{}]{\label{231411}
\includegraphics[width=0.45\textwidth]{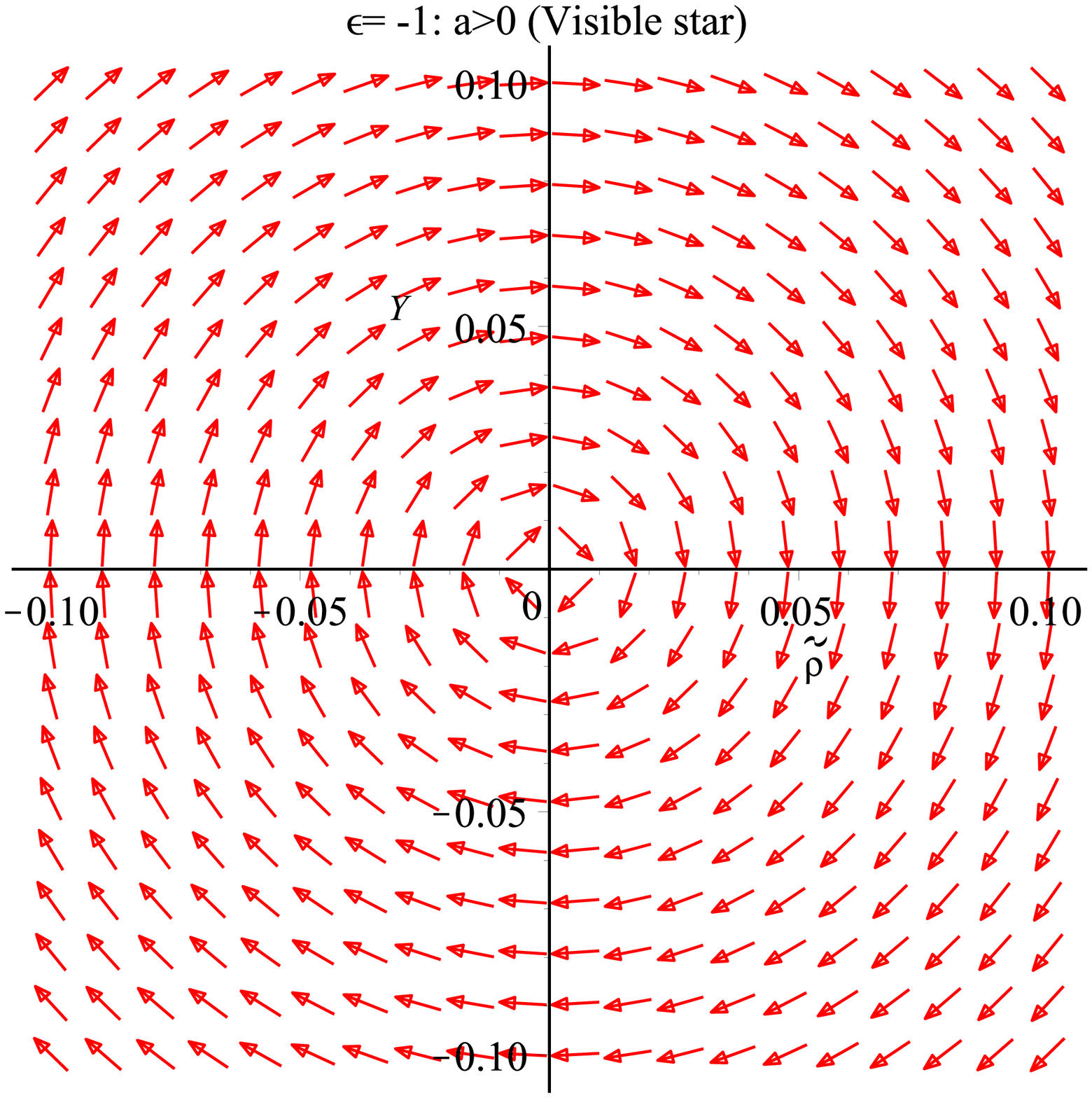}}
\hspace{2mm}\subfigure[{}]{\label{23s1411}
\includegraphics[width=0.45\textwidth]{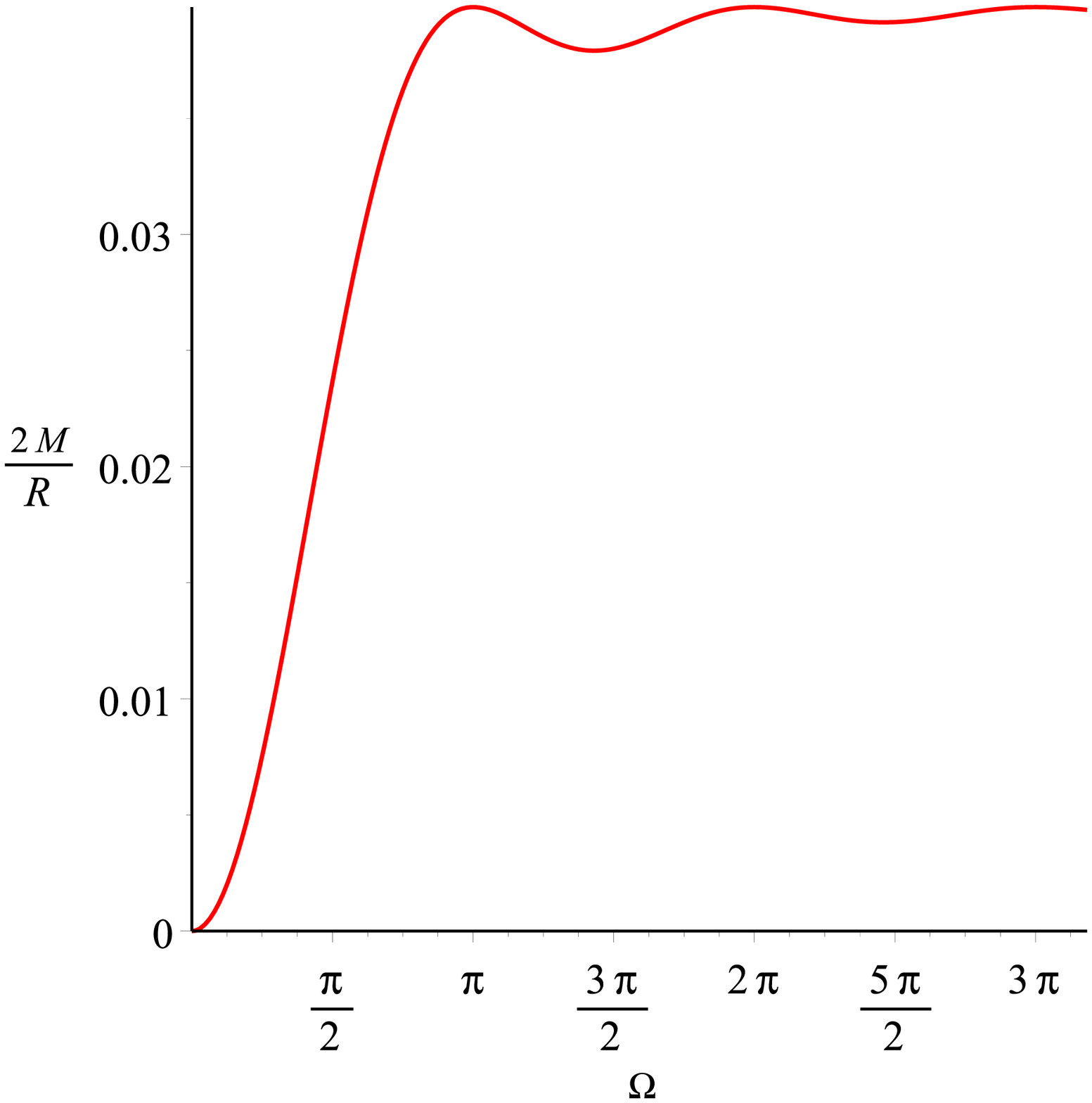}}
\hspace{2mm}\subfigure[{}]{\label{23d1411}
\includegraphics[width=0.45\textwidth]{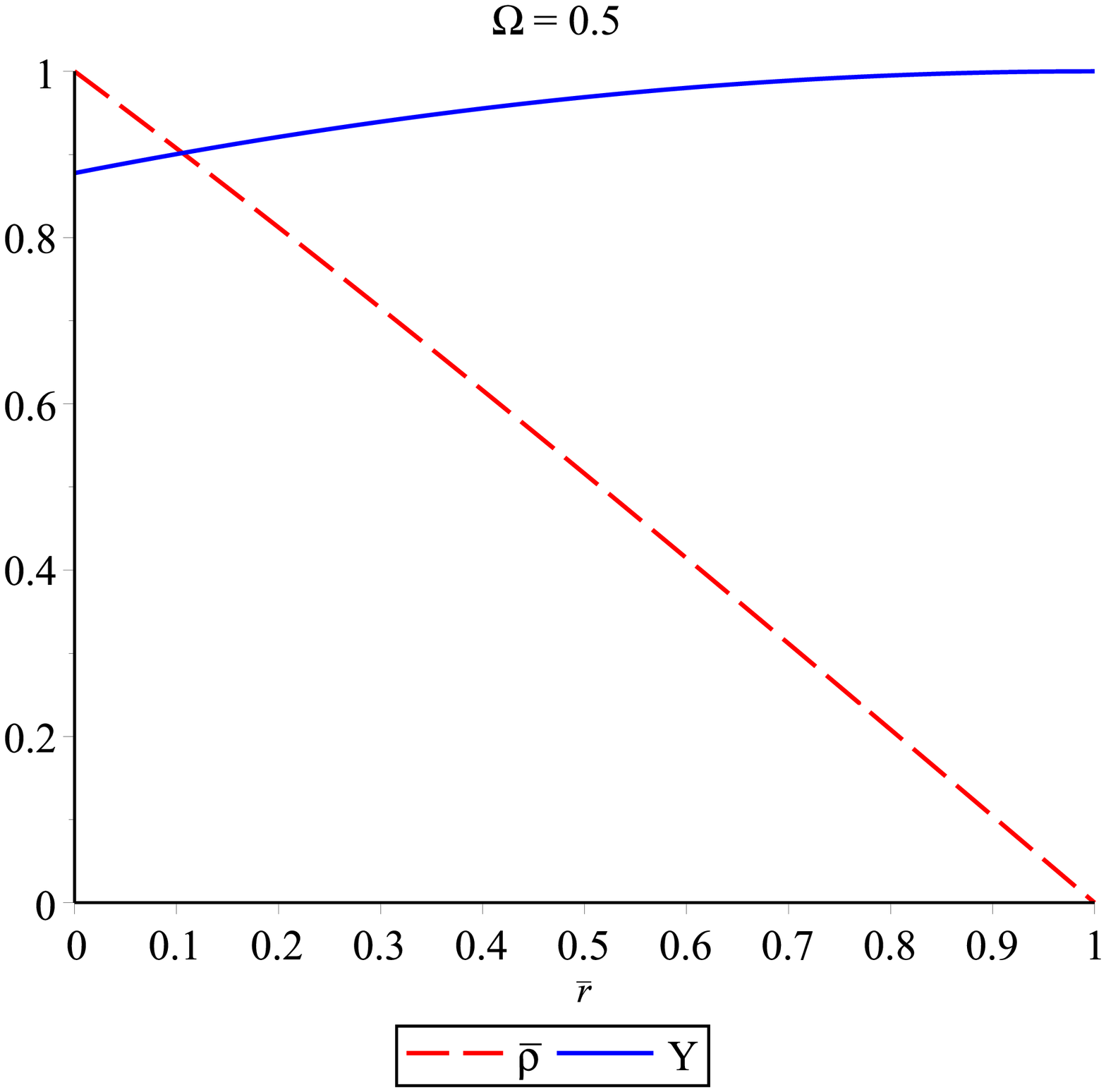}}
\hspace{2mm} \caption{\footnotesize{Arrow diagrams for dark sector
of stellar object (Negative pressure $a<0$) (a) and visible
stellar object (Positive pressure $a>0$) (b), Buchdahl inequality
(compactness) parameter (c), density function $\bar{\rho}$ and
metric field $Y$ for $\Omega=0.5$ (d)}}
\end{figure}

\begin{figure}
\centering\subfigure[{}]{\label{1011}
\includegraphics[width=0.45\textwidth]{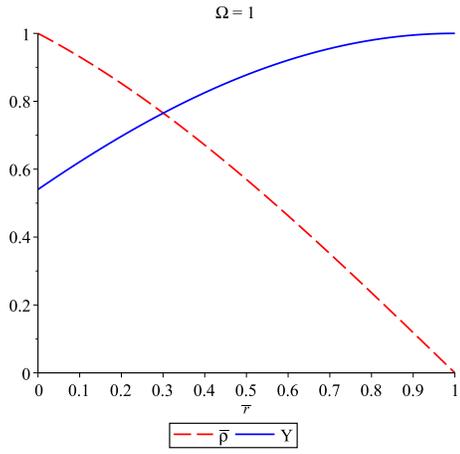}}
\hspace{2mm}\subfigure[{}]{\label{23411}
\includegraphics[width=0.45\textwidth]{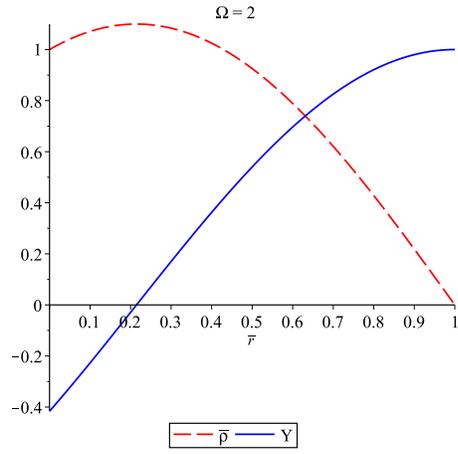}}
\hspace{2mm}\subfigure[{}]{\label{23s411}
\includegraphics[width=0.45\textwidth]{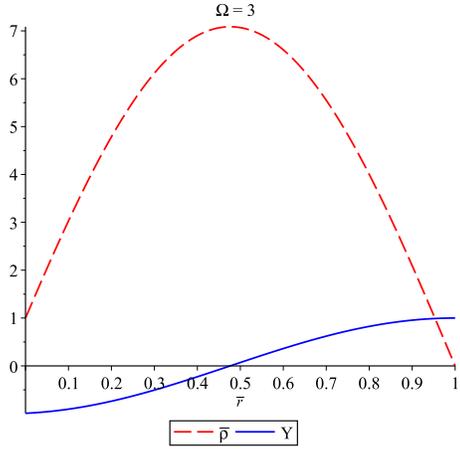}}
\hspace{2mm}\subfigure[{}]{\label{23d411}
\includegraphics[width=0.45\textwidth]{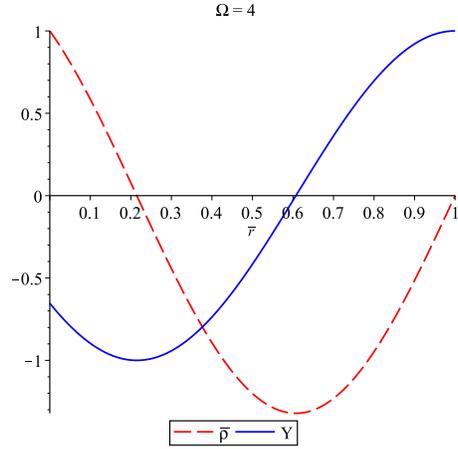}}
\hspace{2mm} \caption{\footnotesize{Diagrams of energy density and
metric field for $\Omega=1$ (a), $\Omega=2$ (b), $\Omega=3$ (c)
and $\Omega=4$ (d)}}
\end{figure}
\begin{figure}
\centering\subfigure[{}]{\label{1013}
\includegraphics[width=0.33\textwidth]{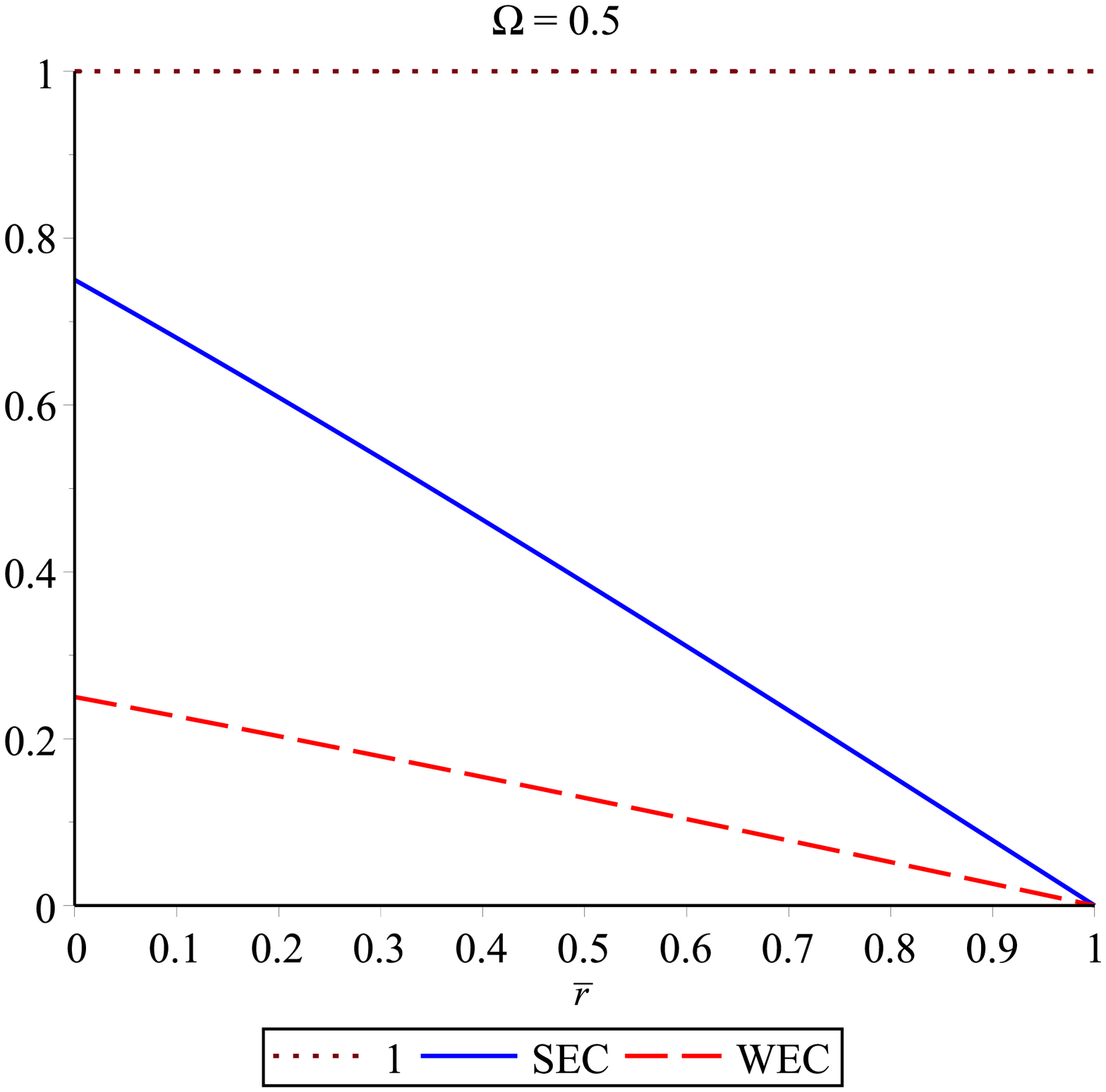}}
\hspace{2mm}\subfigure[{}]{\label{s111}
\includegraphics[width=0.33\textwidth]{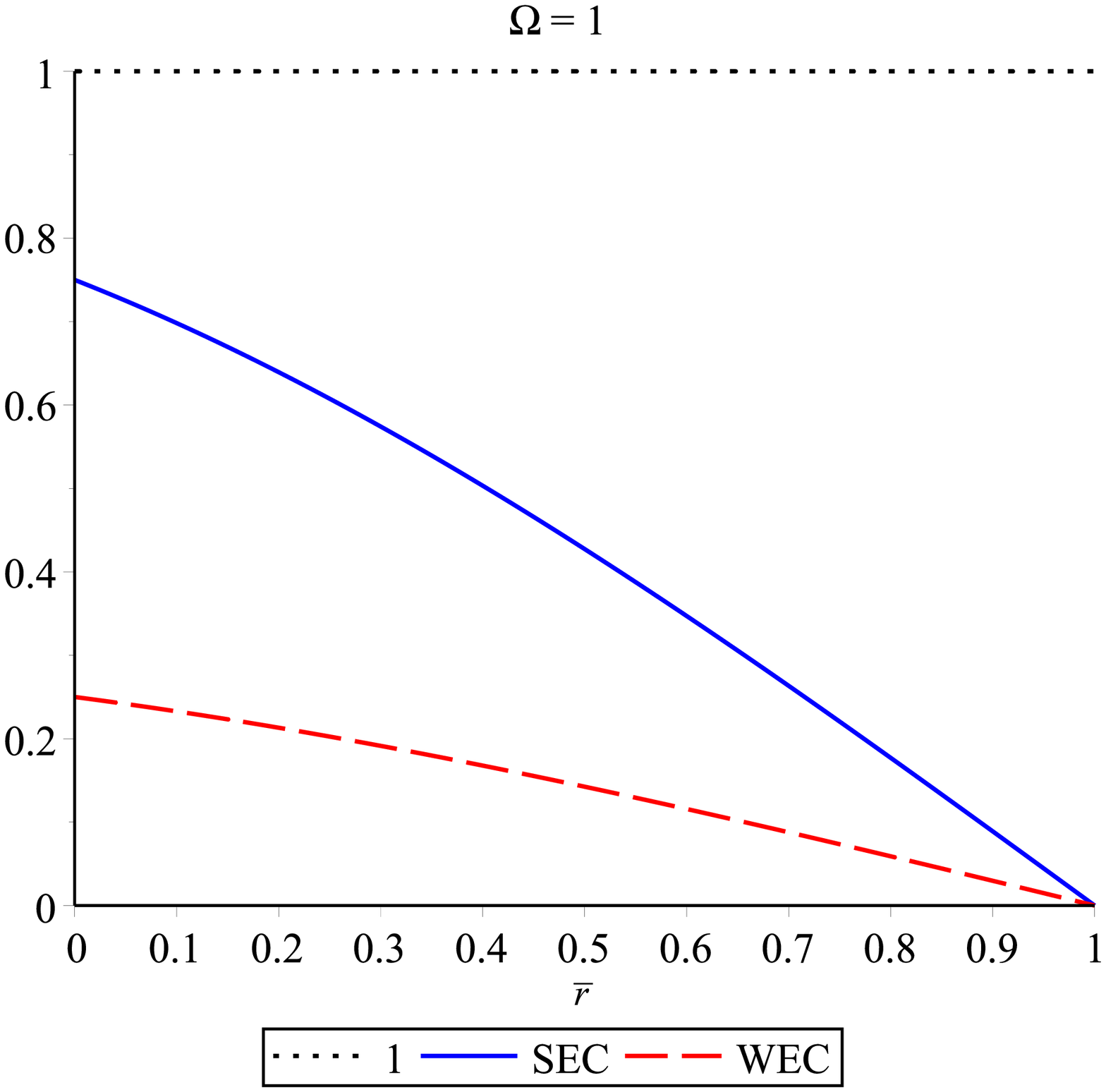}}
\hspace{2mm}\subfigure[{}]{\label{eifgen}
\includegraphics[width=0.33\textwidth]{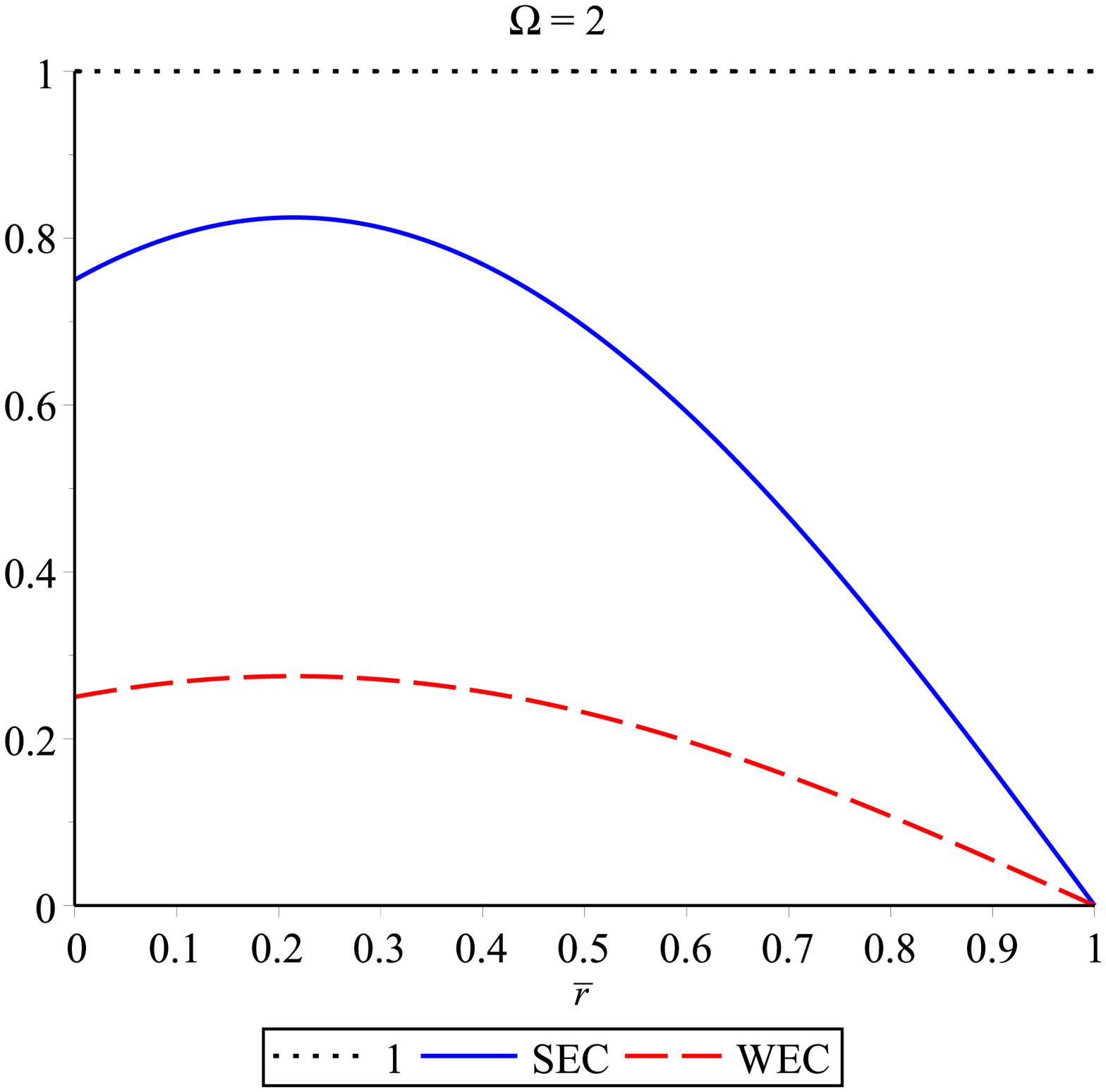}}
\hspace{2mm}\subfigure[{}]{\label{qq}
\includegraphics[width=0.33\textwidth]{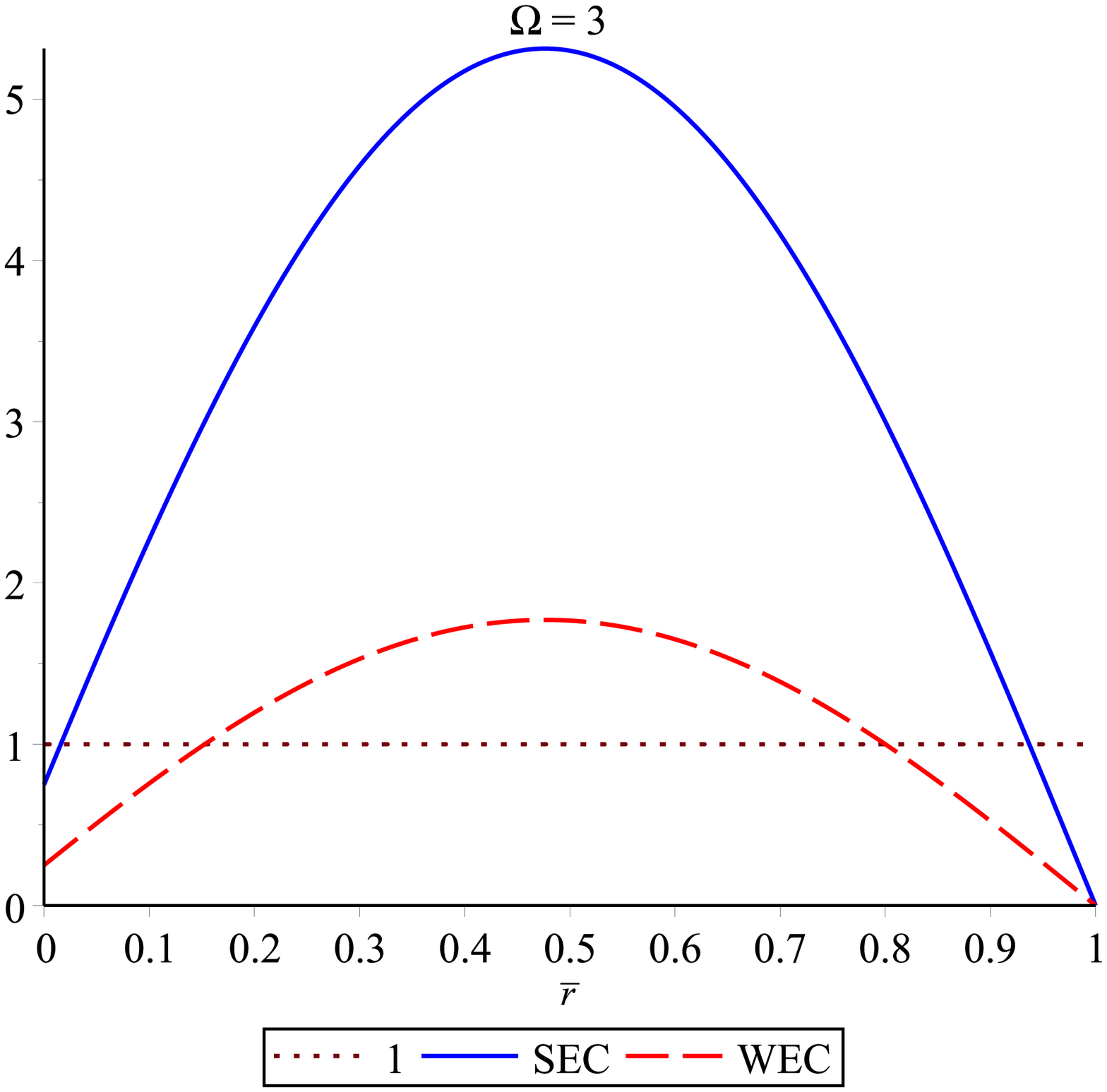}}
\hspace{2mm}\subfigure[{}]{\label{qq}
\includegraphics[width=0.33\textwidth]{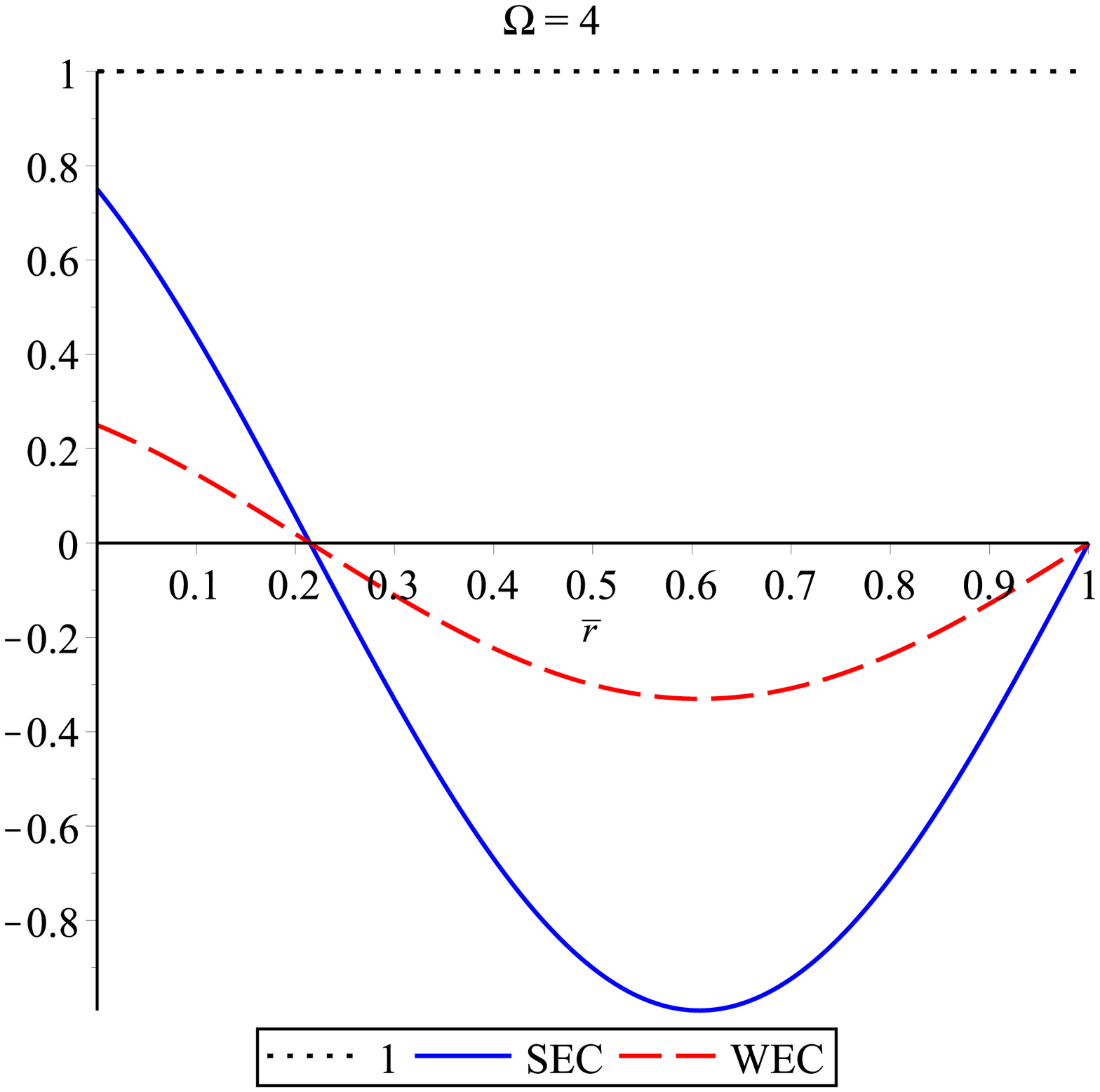}}
\hspace{2mm}\caption{\footnotesize{Diagrams for WEC and SEC
inequalities for different values of the $\Omega$ parameter  }}
\end{figure}
\begin{figure}
\centering\subfigure[{}]{\label{1013}
\includegraphics[width=0.33\textwidth]{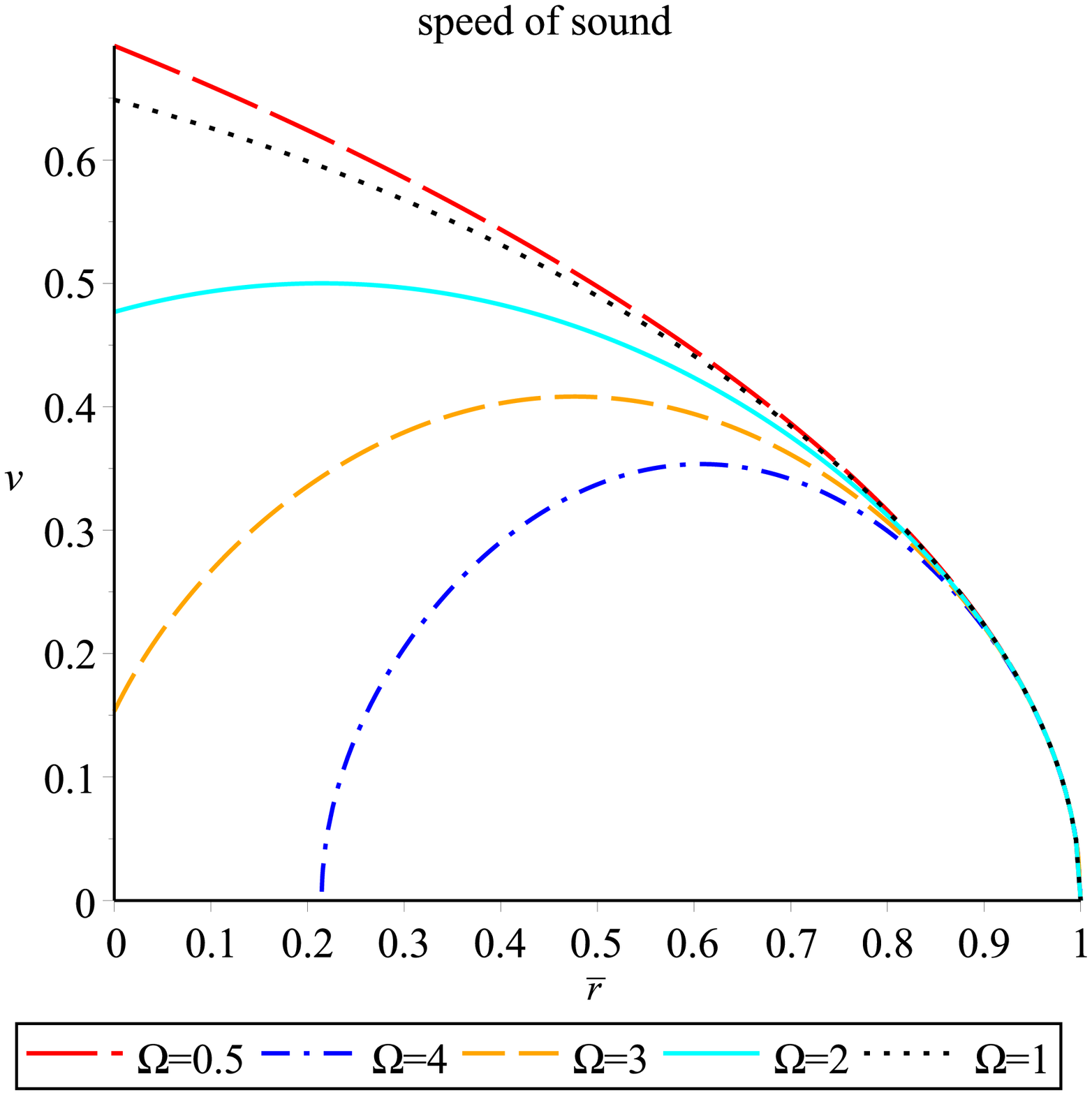}}
\hspace{2mm}\subfigure[{}]{\label{s111}
\includegraphics[width=0.33\textwidth]{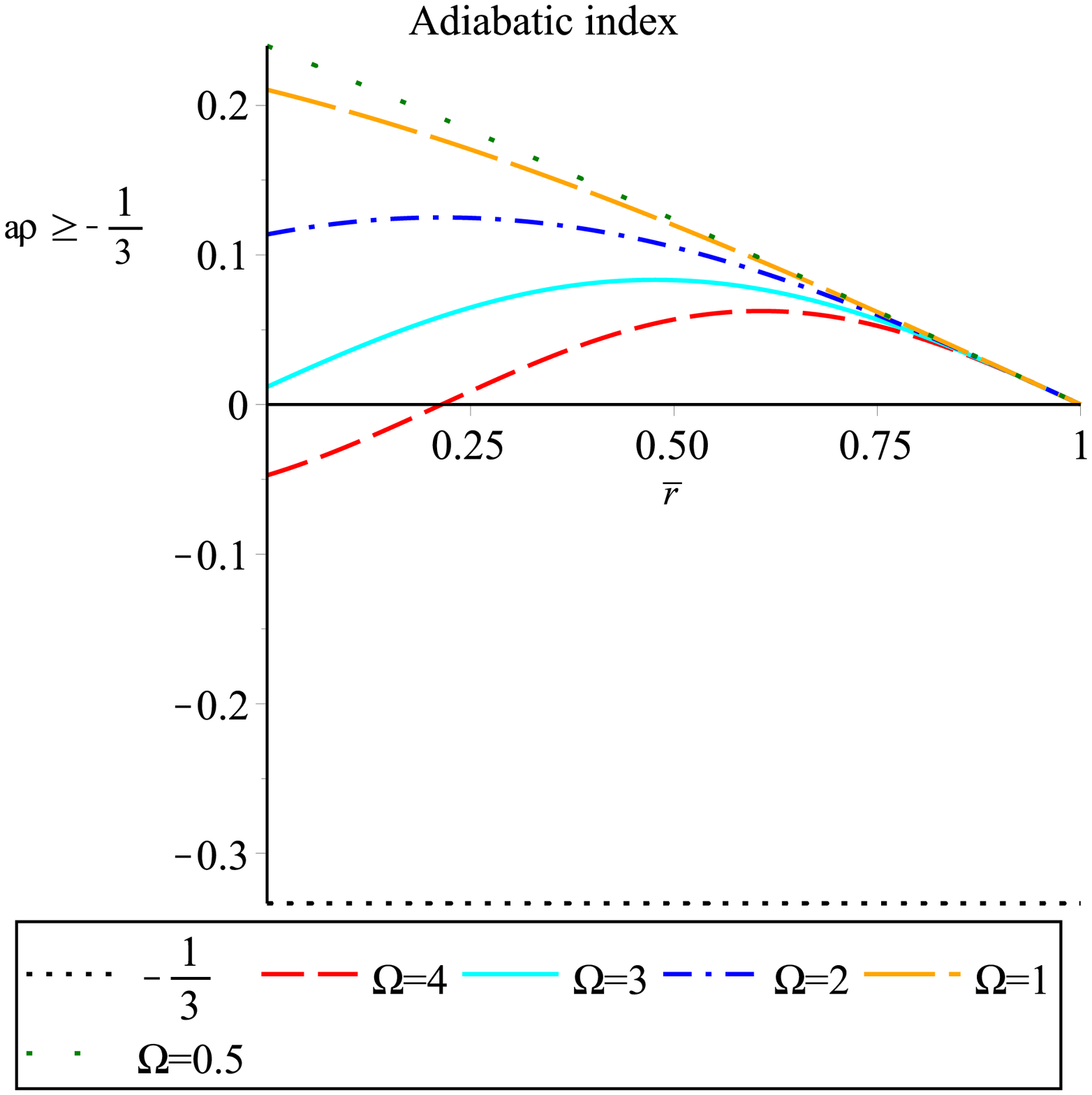}}
\hspace{2mm}\subfigure[{}]{\label{eifgen}
\includegraphics[width=0.33\textwidth]{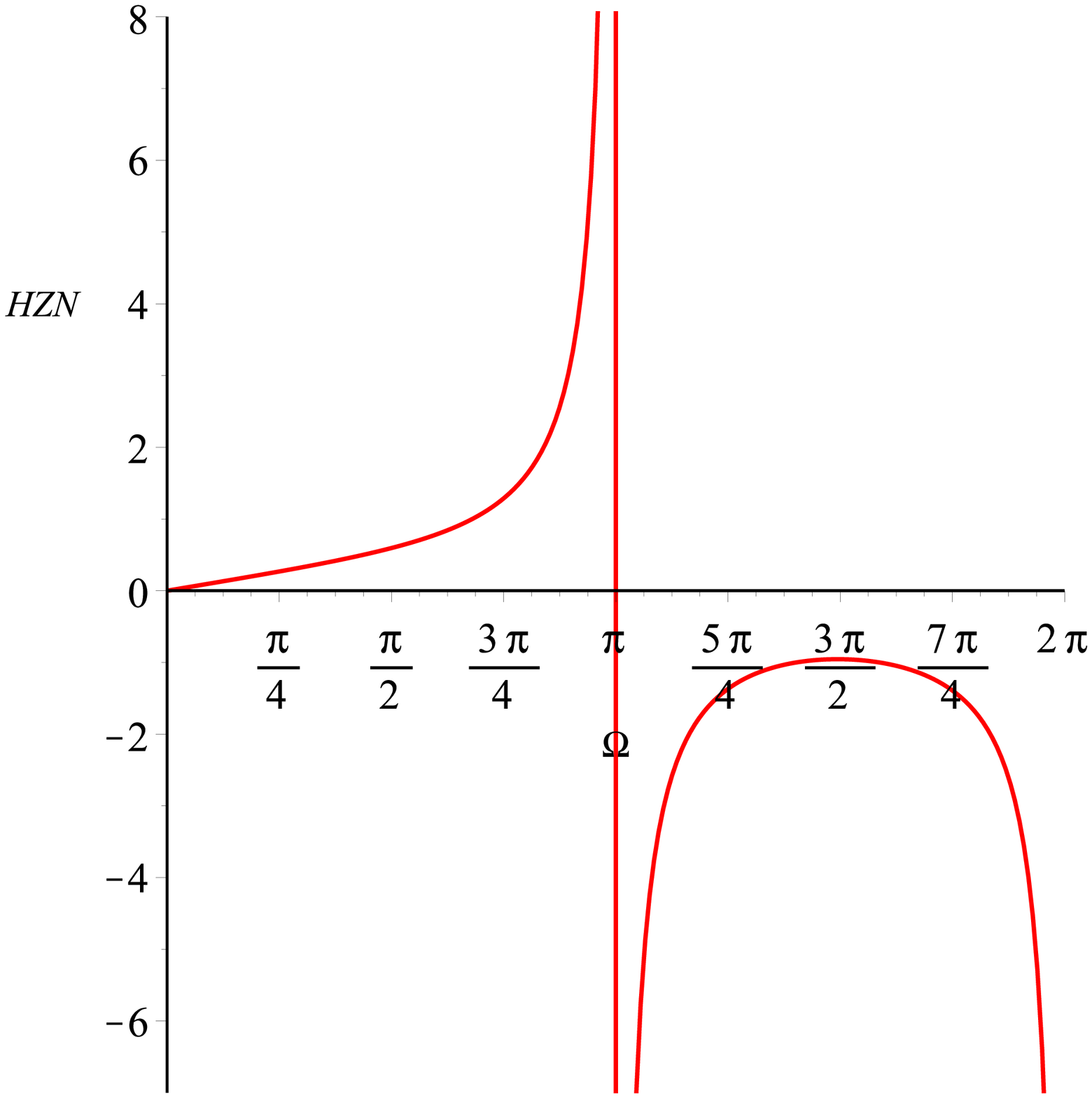}}
\hspace{2mm}\caption{\footnotesize{Diagrams for speed of sound (a)
which is less than the light velocity describing a stable state.
Diagram of adiabatic index (b) for which $a\rho\geq-\frac{1}{3}$
defines stabilization of the system.
 }}
\end{figure}
\end{document}